\documentclass{elsart}
\usepackage{graphicx}
\usepackage{natbib}

\begin{document}

\begin{frontmatter}
   \title{High Efficiency of Gamma-Ray Bursts Revisited}
   \author{Y. C. Zou$^a$ and Z. G. Dai $^b$ }
   \address{$^a$Department of Physics, Huazhong University of Science and Technology, 430074 Wuhan, China \\
                $^b$Department of Astronomy, Nanjing University, 210093 Nanjing, China
           }
   \ead{zouyc@hust.edu.cn, dzg@nju.edu.cn}

   \begin{abstract}
    Using the conservation of energy and momentum during collisions of
    any two shells, we consider the efficiency of gamma-ray bursts
    by assuming that the ejecta from the central engine are equally
    massive and have the same Lorentz factors. We calculate the efficiency
    and the final Lorentz factor of the merged whole shell for different
    initial diversities of Lorentz factors and for different microscopic
    radiative efficiency. As a result, a common high efficiency in the
    range of 0.1 to 0.9 is considerable, and a very high value near 100\%
    is also reachable if the diversity of the Lorentz factors is large enough.
   \end{abstract}
   \begin{keyword}
    {gamma rays: bursts, theory}
   \end{keyword}
\end{frontmatter}

\section{Introduction}
\label{sect:intro}

The efficiency of a gamma-ray burst (GRB) $\eta$ has aroused great
interests in the theoretical studies \citep{Meszaros1993, Xu2004},
because it firmly links with the total energy of the central engine
ejected. The energy of the gamma rays is almost standard energy with
about $1.3\times 10^{51} \rm{erg}$\citep{Bloom2001}. Thus, if the
efficiency is too low, the total ejected energy should be too large
and incredible. If it is too high, however, the required large
velocity diversity of the ejected shells is far from understanding.
Many previous works calculated the efficiency, but the results are
rather different. \citet{Kumar1999} drew the conclusion that the
efficiency can only be around 1\%, while \citet{Beloborodov2000}
argued that it can reach about 100\%.  Considering different models
for the velocity diversities, \citet{Kobayashi2001, Guetta2001}
found that the results are highly model-dependent, and the
efficiency varies in the range of 0.1 to 0.9.

The efficiency can also be obtained by the model fitting for
individual bursts \citep[e.g.][]{Freedman2001, Panaitescu2002}. They
found that the $\eta$ varies for different bursts, and some bursts
have a very high value about 0.9 \citep{Freedman2001}. Some
statistics were carried out for the efficiency by
\citet{Lloyd-Ronning2004, Eichler2005}, who found that it may be
related to some energies such as the peak energy or the total energy
of gamma-rays.

Recently, the shallow decay phase of early X-ray afterglows
\citep{Nousek2006} appeared in many bursts detected by Swift. This
phenomenon is generally interpreted as post-burst energy injection.
This model enhances the efficiency enormously \citep{Eichler2005}.
\citet{Granot2006} suggested alternate models (e.g., two-component
jet model) to avoid this paradox, while by considering the reverse
Compton scattering effect, \citet{Fan2006} found that the efficiency
may not be too large even with the energy injection.

In the dynamics, the essential problem is the diversity of the
velocities (or Lorentz factors) of sub-shells ejected from the
central engine. It is not natural for the central engine to eject
shells with very high  diversity. Enlightened by the equal energetic
ejecta coming from a differentially rotating pulsar \citep{Dai2006},
we suggest that the ejected shells from the central engine are all
equal, and the burst itself and the followed X-ray
flares\citep{Nousek2006} all originate from collisions between
ejected shells and a slow proceeding shell at the front, which may
originally be the envelope of the massive progenitor and be
accelerated by the rapid overtaken shells. This model has been used
to successfully fit the multi-band observations of the burst GRB
050904\citep{Zou2006}. In this model, the initial Lorentz factor of
the envelope may be unity at the beginning. In this paper, we
consider the dynamical evolution of the envelope and the collisions
with the upcoming ejected shells, and calculate the emission
efficiency, which is taken as the efficiency of GRBs. In this model,
the emitted energy of photons can be much greater than the final
kinetic energy. We give the dynamics in section
\S\ref{sec:analysis}, calculate the efficiency numerically in
\S\ref{sec:numeric}, and summarize our results in
\S\ref{sec:conclusion}.

\section{Analysis} \label{sec:analysis}
We consider one collision between two shells with mass $m_1$ and
$m_2$, Lorentz factor $\gamma_1$ and $\gamma_2$ respectively. They
merge into one whole shell with mass $m$ (where the materials may be
hot) and Lorentz factor $\gamma$. We assume that a fraction
$\epsilon$ of the internal energy will convert into photons. This
process should obey the conservation of energy and momentum:
\begin{eqnarray}
   \gamma_1 m_1 c^2 + \gamma_2 m_2 c^2 = \gamma m c^2 + E_{\gamma}, \label{eq:energy_con}\\
   \gamma_1 m_1 \beta_1 c +\gamma_2 m_2 \beta_2 c = \gamma m \beta c +E_{\gamma} /c, \label{eq:momentum_con}
\end{eqnarray}
where $\beta$ is the velocity in units of $c$ (corresponding to the
Lorentz factor $\gamma$), $E_{\gamma} = \gamma \epsilon E$ is
radiated as photons and $(1-\epsilon)E$ remains in the merged shell,
while $E$ is the additional internal energy produced in this
collision in the comoving frame of the merged shell. The mass of the
merged shell is $m=m_1 +m_2 +(1-\epsilon)E/c^2$. Define
$\epsilon$ as the radiation efficiency, which should be
differentiated with the total efficiency of the gamma-ray burst
$\eta$. For one collision, the efficiency is $\eta =
E_{\gamma}/(\gamma_1 m_1 c^2+\gamma_2 m_2 c^2)$. For $M$ shells with
$N$ collisions, the total efficiency is then
\begin{equation}
   \eta = \frac{\sum_{i=1}^N E_{\gamma,(i)}}{\sum_{i=1}^{M}\gamma_{(i)} m_{(i)} c^2}.
   \label{eq:effi}
\end{equation}

From these equations, one can find that the evolution depends very
sensitively on the collision history. For example, if all the rapid
shells are merged first, which cannot produce many photons for a low
diversity of velocities, and then, the whole merged shell overtakes
a slowly proceeding shell, then the final efficiency will be very
small, provided that the slow shell has much less mass than the
whole fast shell. A more efficient strategy is that faster shells
catch up with slower ones one by one. In this model, all the nearly
equal energetic and massive shells colliding with the envelope of
the progenitor\citep[like][]{Zou2006}, it just satisfies the
efficient strategy. We can expect that a high efficiency appears in
this scenario. As all the shells are colliding with the sole
foregoing one, we obtain $M=N+1$. However, to get the
final efficiency, Eqs. (\ref{eq:energy_con}) and
(\ref{eq:momentum_con}) should be used in each collision, numerical
simulations must be performed.

\section{Calculations} \label{sec:numeric}
We assume: $\gamma_1$ and $m_1$ are the initial Lorentz factor and
mass of the slow shell in the front;
$\gamma_2$ and $m_2$ are the Lorentz factor and mass of the ejected
shells from the central engine, which are taken as constant in the
model; $\gamma$ is Lorentz factor of the accelerating merged shell,
which is originally $\gamma_1$, and afterwards, mixed with the
ejected shells. We suppose that the ejected shell with mass $m_2$
and Lorentz factor $\gamma_2$ collides with the envelope. They merge
into one shell proceeding with Lorentz factor $\gamma$, and then
another ejected shell with same mass $m_2$ and Lorentz factor
$\gamma_2$ collides with the merged shell, which proceeds with a new
Lorentz factor $\gamma$, and so on.

\begin{figure}
  \includegraphics[width=1.\textwidth,angle=270]{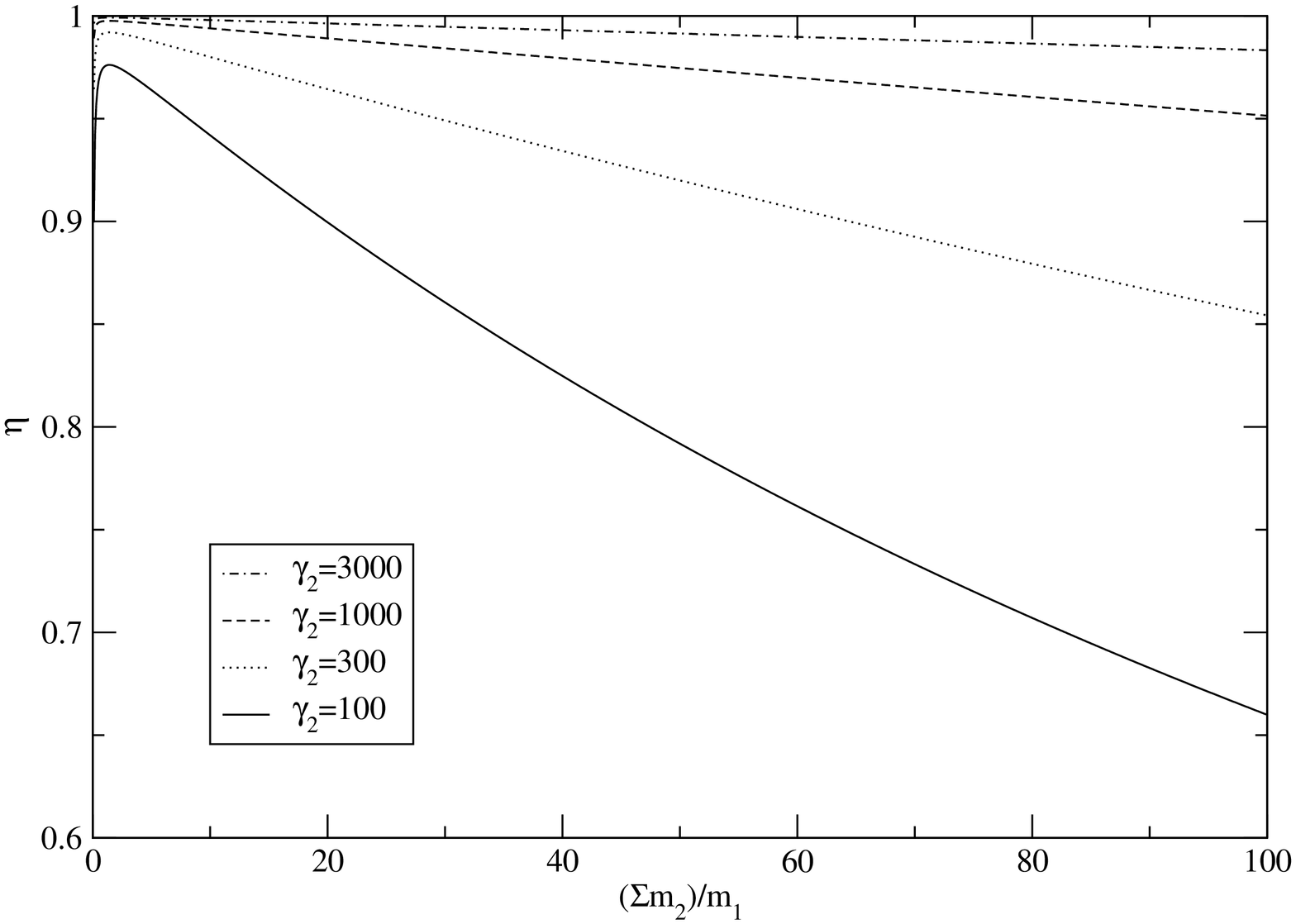}
  \caption{The efficiency of a gamma-ray burst as a function of total ejected mass over
   $m_1$ ($(\Sigma m_2)/m_1$). The Lorentz factor of the foregoing slower shell is
   set as $\gamma_1=1$, and the radiative efficiency for each collision is $\epsilon=1$.
   The different curves correspond to different Lorentz factor of the ejected fast shells,
   which are marked in the figure.}
  \label{fig:ef_eps-1_g1-1}
\end{figure}

\begin{figure}
  \includegraphics[width=1.\textwidth,angle=270]{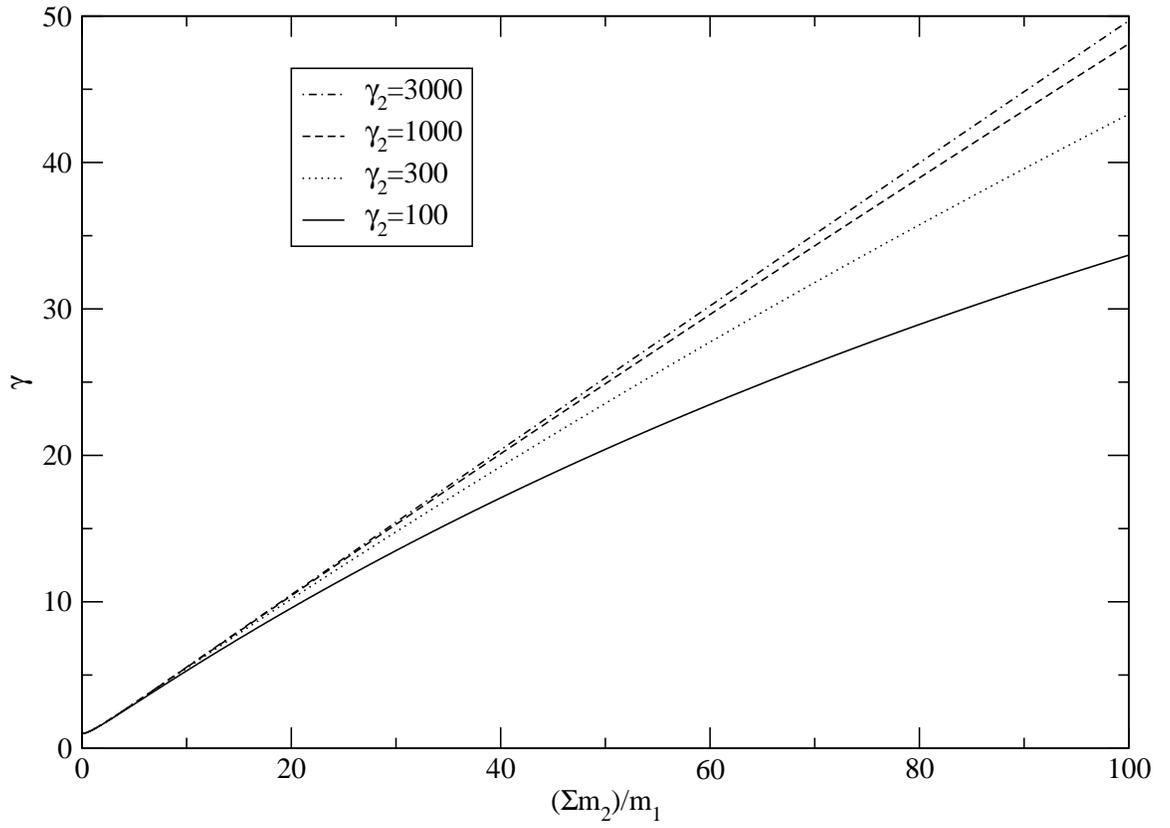}
  \caption{The Lorentz factor of the merged shell as function of $(\Sigma m_2)/m_1$,
  with $\gamma_1=1$ and $\epsilon=1$. Different curves are for the same cases as in
  Fig. \ref{fig:ef_eps-1_g1-1}.}
  \label{fig:g_eps-1_g1-1}
\end{figure}

Fig. \ref{fig:ef_eps-1_g1-1} shows the efficiency varying with the
total mass of ejected shells.  
The efficiency increases with $(\Sigma m_2)/m_1$ first,
because when the $m_1 > \Sigma m_2$, from Eq. \ref{eq:effi}, 
the denominator can be regarded as constant approximately, while
the numerator increase with the number of collisions. With the
increase of $(\Sigma m_2)/m_1$, the Lorentz factor of merged shell
$\gamma$ approach the Lorentz factor of the faster shells $\gamma_2$
more and more closely, then the collisions produce less and 
less emissions, therefore, the efficiency decreases.
Another clear tendency is that, for larger
value of $(\Sigma m_2)/m_1$, the collisions are less efficient but
with higher final Lorentz factor, and for smaller value of  $(\Sigma
m_2)/m_1$, the collisions are more efficient but with lower Lorentz
factor. 
For the case $\gamma_2=1000$, the efficiency is greater than
0.95, even though the total ejected shells is 100 times more massive
than the envelope. At the same time, one should consider the final
Lorentz factor to be reasonable. As this final Lorentz factor is the
initial Lorentz factor for the afterglow, the value of final Lorentz
factor should be in the range of tens to hundreds. Fig.
\ref{fig:g_eps-1_g1-1} shows the final Lorentz factor corresponding
to the cases shown in Fig. \ref{fig:ef_eps-1_g1-1}. It shows, for
these values of $\gamma_2$, $(\Sigma m_2)/m_1$ should not be less
than 20, and a value larger than 100 is also reasonable. For larger
$\gamma_2$, the efficiency and the final Lorentz factor become
larger. But a very large $\gamma_2$ may be difficult to occur from
the central engine.

These figures may have one implication about the evolution history
of the merged shell. The merged shell is initially the envelope at
rest. It is accelerated with the accumulation of ejected fast
shells. The efficiency increases before the $(\Sigma m_2)$ is less
than several times $m_1$, and then it decreases.

\begin{figure}
  \includegraphics[width=1.\textwidth,angle=270]{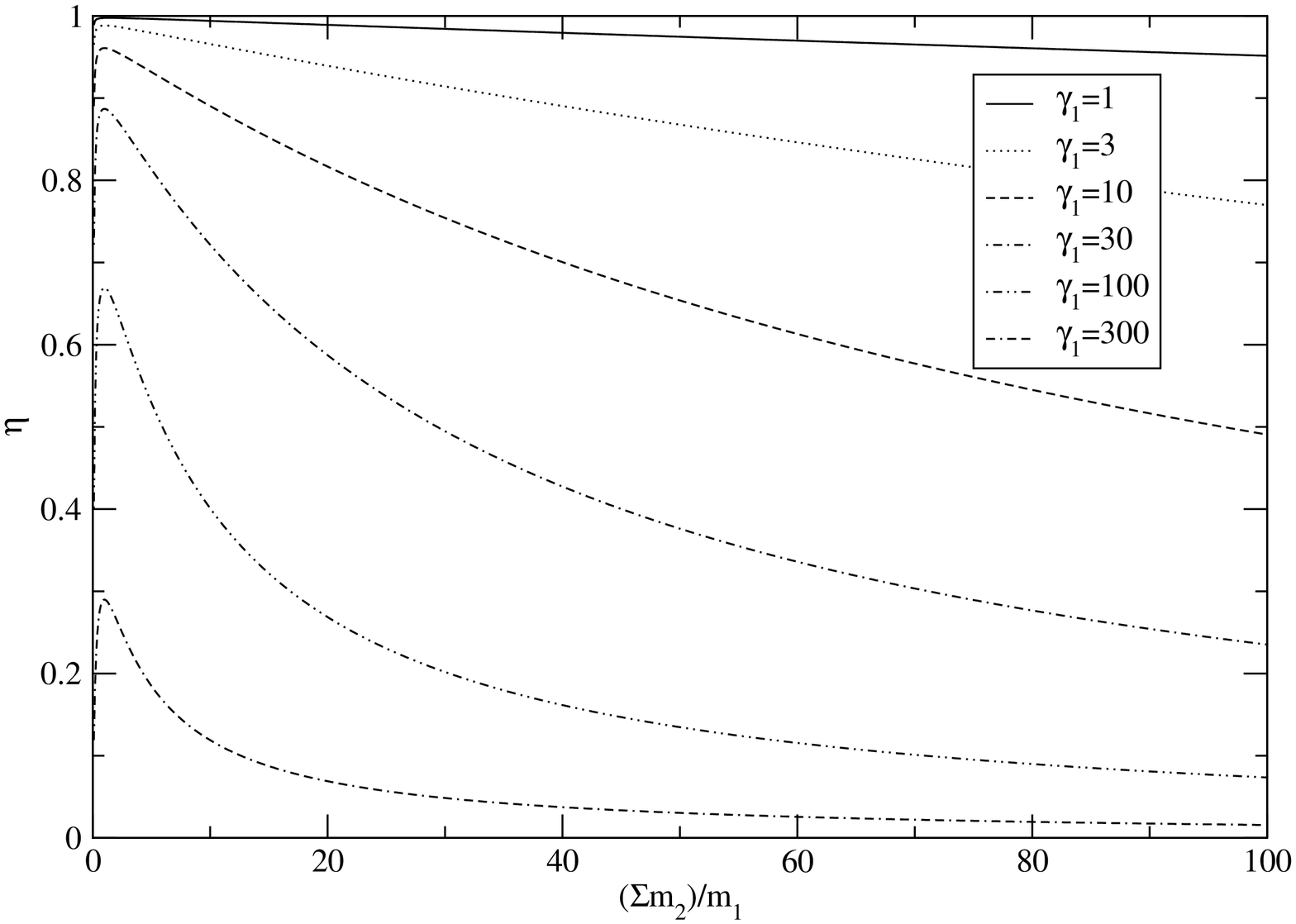}
  \caption{Same as Fig. \ref{fig:ef_eps-1_g1-1} with parameters $\gamma_2=1000$,
  $\epsilon=1$ and different $\gamma_1$.}
  \label{fig:ef_eps-1_g2-1000}
\end{figure}

\begin{figure}
  \includegraphics[width=1.\textwidth,angle=270]{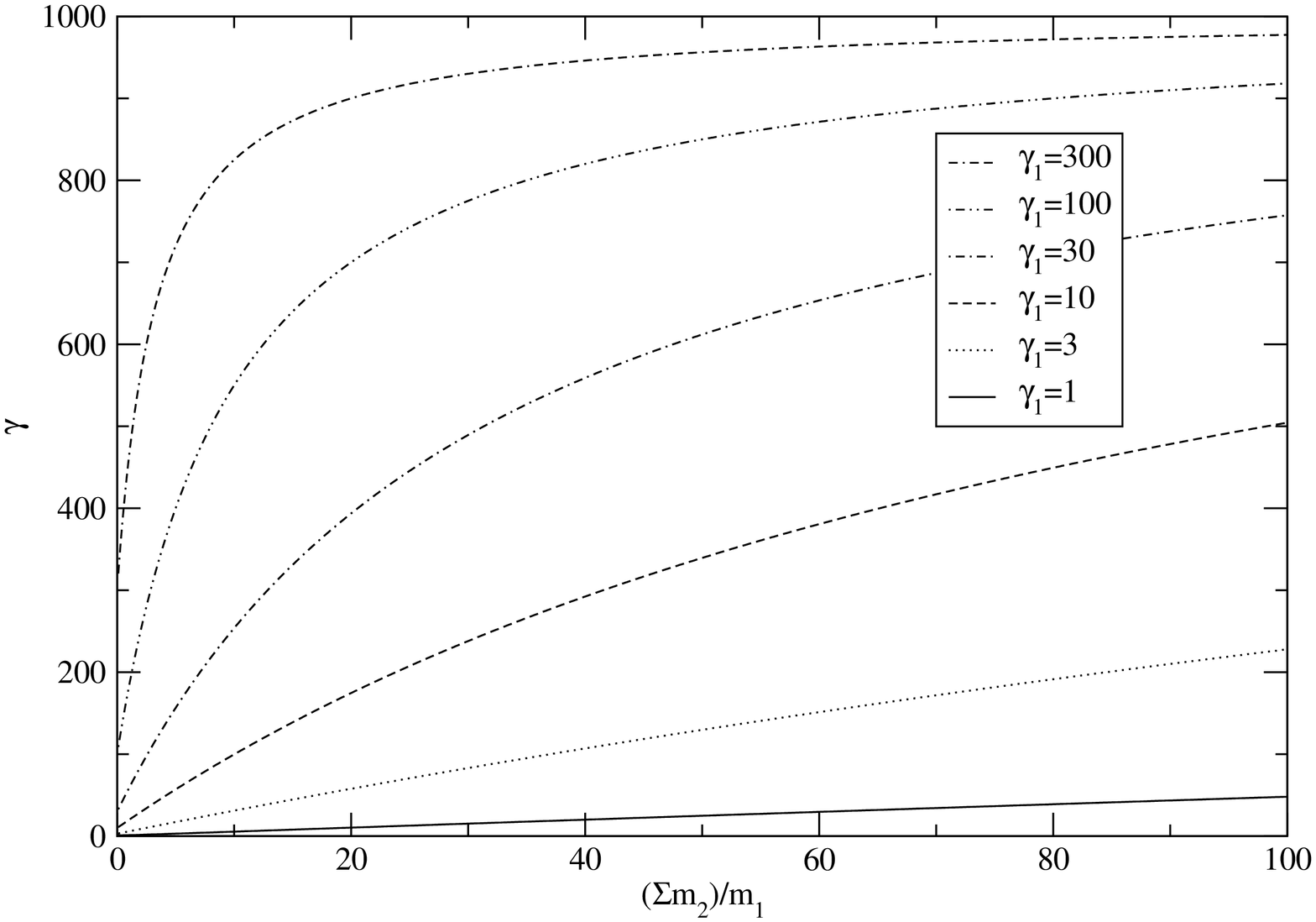}
  \caption{Same as Fig. \ref{fig:g_eps-1_g1-1} with parameters $\gamma_2=1000$,
  $\epsilon=1$ and different $\gamma_1$.}
  \label{fig:g_eps-1_g2-1000}
\end{figure}

It is possible that the radius of the envelope is too small, and the
number density of electrons is too large, which prevents the
gamma-rays to radiate because of a high optical depth. Therefore,
the envelope may be accelerated by the collisions but few photons
radiate at first. With the collision radius increasing, the merged
shell becomes optically thin. We can simply consider this optical
thin shell as the initial foregoing shell, with mass $m_1$ and
$\gamma_1>1$. We calculate the efficiency and the final Lorentz
factor for different $\gamma_1$, while $\epsilon$ and $\gamma_2$ are
constant, which are shown in Figs. \ref{fig:ef_eps-1_g2-1000} and
\ref{fig:g_eps-1_g2-1000}. Generally, the efficiency is less and the
final Lorentz factor is larger than those in the case of
$\gamma_1=1$, which are also plotted in Figs.
\ref{fig:ef_eps-1_g2-1000} and \ref{fig:g_eps-1_g2-1000} as solid
lines. For the case $\gamma_1=300$, the efficiency is too small and
the final Lorentz factor is too large, which should be ruled out for
a normal gamma-ray burst. One may select a set of eclectic
parameters for a real burst.

\begin{figure}
  \includegraphics[width=1.\textwidth,angle=270]{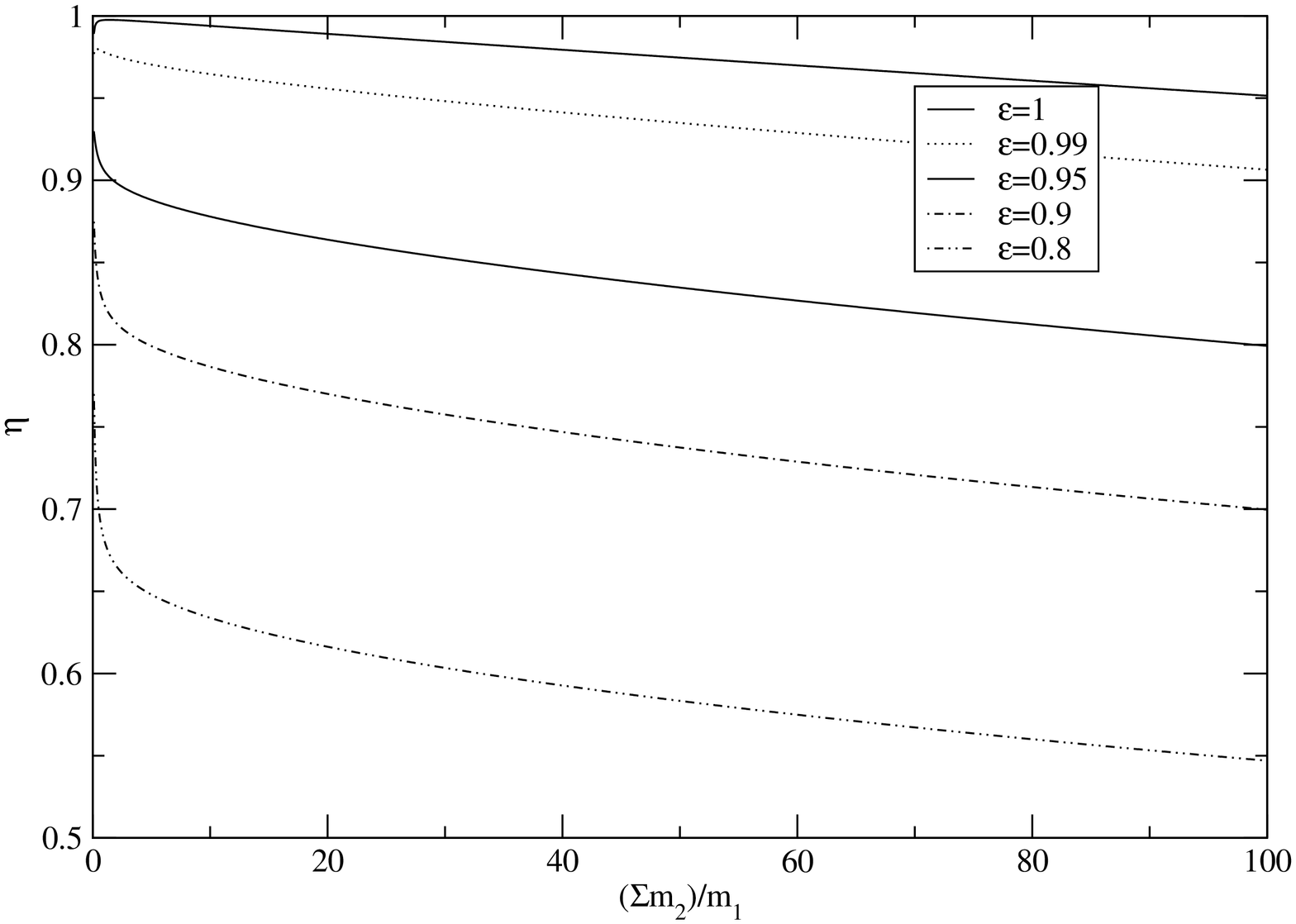}
  \caption{Same as Fig. \ref{fig:ef_eps-1_g1-1} with parameters $\gamma_1=1$,
  $\gamma_2=1000$, and different $\epsilon$.}
  \label{fig:ef_g1-1_g2-1000}
\end{figure}

\begin{figure}
  \includegraphics[width=1.\textwidth,angle=270]{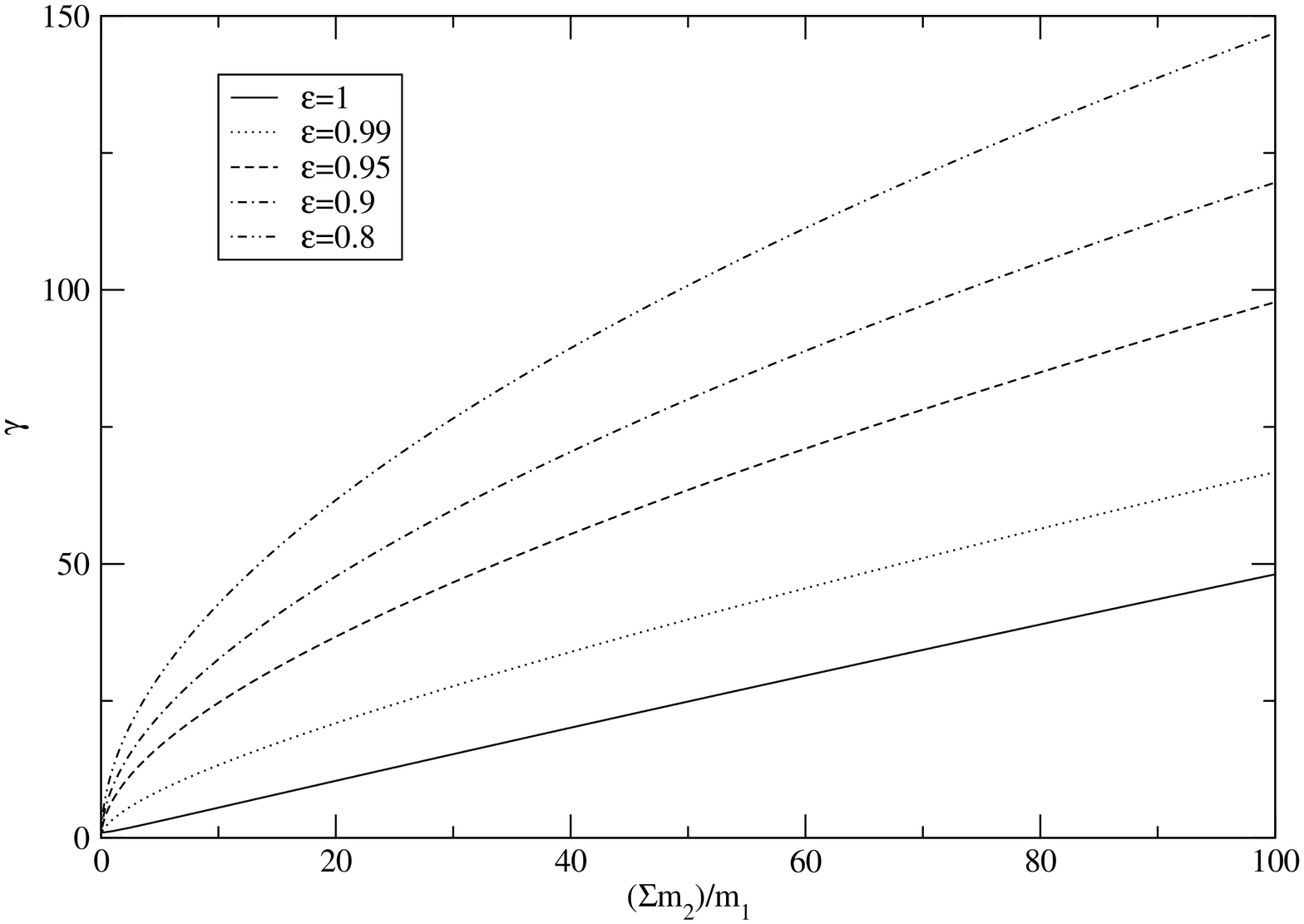}
  \caption{Same as Fig. \ref{fig:g_eps-1_g1-1} with parameters $\gamma_1=1$,
  $\gamma_2=1000$, and different $\epsilon$.}
  \label{fig:g_g1-1_g2-1000}
\end{figure}

The microscopic radiative efficiency $\epsilon$ may not be 1
perfectly, i.e., all the internal energy cannot be transferred into
radiation. But $\epsilon$ shouldn't be too small neither, otherwise,
the internal energy will accumulate more and more with collisions
going on, and then the very high proportion of internal energy will
be emitted out definitely. The efficiency and the Lorentz factor as
function of $(\Sigma m_2)/m_1$ for different $\epsilon$ are plotted
in Figs. \ref{fig:ef_g1-1_g2-1000} and \ref{fig:g_g1-1_g2-1000},
with $\gamma_1=1$ and $\gamma_2=1000$. For simplicity, we set
$\epsilon$ as constant. As a part of the internal energy is not
radiated, the efficiency decreases greatly and the final
Lorentz factor increases appreciably as $\epsilon$ decreases.
In Fig. \ref{fig:ef_g1-1_g2-1000}, for smaller $\epsilon$s, it looks like
the efficiency decreases with the increase of $(\Sigma m_2)/m_1$
directly. In fact, it is just that the stage of increasing stops for less
$(\Sigma m_2)/m_1$. As for smaller $\epsilon$, it means more 
produced internal energy is left in the merged shell. This makes
the total mass of merged shell be comparable with $m_1$ more 
early (see Eq. \ref{eq:effi}), and correspondingly, the decreasing 
stage comes more early.

In the calculations, we assume that $m_2=m_1/10$. But the testing
calculations for different $m_2$ show that the results do not depend
on the particular value of $m_2$, provided $m_2 < m_1$. Therefore,
the mass of ejected shell is not required to be equal for the
results to be valid.

\section{Conclusions}
\label{sec:conclusion}

Considering the conservations of energy and momentum, we calculate
the efficiency of a gamma-ray burst in different cases, assuming
that the ejected shells from the central engine are equally
energetic and massive and they all collide onto a slower shell
proceeding in the front. A general conclusion is that for a large
diversity of the low and high initial Lorentz factors, the
efficiency of bursts is higher, which is consistent with the above
analysis. We have detailedly considered the influences of the values
of initial Lorentz factors and the microscopic radiative efficiency.
The final Lorentz factor depends sensitively on the initial Lorentz
factor of the slower shell, and the efficiency depends more
sensitively on the values of the Lorentz factor of the ejected
shells, while both are sensitive to the microscopic radiative
efficiency. The scenario provided here is a possible solution for
the very high efficiency obtained in the model fittings. Please note
that because the envelopes only exist in collapsars, these
calculations are only suitable for long duration bursts.

In this scenario, there are four free parameters: the initial
Lorentz factor of the slow shell $\gamma_1$ and of the fast shell
$\gamma_2$, the radiative microscopic efficiency $\epsilon$, and the
mass ratio of fast shells and the slow shell $(\Sigma m_2)/m_1$.
These can make the final efficiency be adaptive for a relative large
range, say (0.1,0.9). On the other hand, these parameters will be
restricted by other aspects of observations, e.g, the peak energy,
the deductive Lorentz factor from the afterglow observations, and
son on.

In the view of efficiency, this scenario is the most efficient one to produce
gamma-rays. In other scenarios, the efficiency is relatively low.
For example, if all the faster shells
merge first, which will not produce any photons because of the same
Lorentz factors, and then the merged shell collides with the foregoing slow one, 
the emission will be very inefficient.

\section*{Acknowledgements}
The authors thank Y. F. Huang, X. F. Wu and L. Shao for helpful
discussions. This work was supported by the Natural Science
Foundation of China (grants 10233010 and 10221001).

\newpage

\label{lastpage}

\end{document}